\documentclass[pra,twocolumn,showpacs,superscriptaddress,groupedaddress]{revtex4-1}

\usepackage{amsfonts,hyperref}
\usepackage{graphicx}
\usepackage{amssymb}
\usepackage{epsfig}
\usepackage{amsmath}
\usepackage{xcolor}

\begin{document}

\title{Gap solitons of super-Tonks-Girardeau gas in one-dimensional periodic potential}

\author{T. F. Xu$^{1}$, X. L. Jing$^{1}$, H. G. Luo$^{2,3}$, and C. S. Liu$^{1}$}

\affiliation{$^1$Department of Physics, Yanshan University, Qinhuangdao 066004, China\\
$^2$Center for Interdisciplinary Studies and Key Laboratory for Magnetism and
Magnetic Materials of the MoE,
Lanzhou University, Lanzhou 730000, China\\
$^3$Beijing Computational Science Research Center, Beijing 100084, China}

\begin{abstract}
We study the stability of gap solitons of the super-Tonks-Girardeau bosonic gas in one-dimensional periodic potential. The linear stability analysis indicates that increasing the amplitude of periodic potential
or decreasing the nonlinear interactions, the unstable gap solitons can become stable. In particular, the theoretical analysis and numerical
calculations show that, comparing to the lower-family of gap solitons, the higher-family of gap solitons are
easy to form near the bottoms of the linear Bloch band gaps. The numerical
results also verify that the composition relations between various gap solitons and nonlinear Bloch waves are general and can exist in the super-Tonks-Girardeau phase.
\end{abstract}

\pacs{03.75.Lm, 67.85.-d}
\maketitle


\section{Introduction}

The recent development of trapping and cooling techniques enable
experimental realizations of low-dimensional ultracold atomic and molecular
gases in optical lattices. In particular, by the magnetic Feshbach resonance
or confinement-induced resonance methods, the low-dimension particles can be adjusted
continually from strong repulsive to attractive interaction. It is therefore possible to realize strongly correlated
low-dimensional quantum systems experimentally \cite{RevModPhys.78.179}.

For the weakly repulsive regime in one dimension (1D), the degenerate Bose
gas acts as a quasi Bose Einstein Condensation (BEC). As the strength of the
repulsive interaction tends to infinity, the bosons behave like impenetrable
fermions and the system is known as the Tonks-Girardeau (TG) gas \cite%
{PhysRev.50.955, JMP.1.516}. On the contrary, for the strongly attractive regime, the
1D Bose gas is more strongly correlated than the TG gas and can be stable in
a wide range of strongly attractive interaction strength, and thus called it
as the super Tonks-Girardeau gas (STG) phase \cite{PhysRevLett.95.190407,
1742-5468-2005-10-L10001, 1367-2630-10-10-103021}. The STG gas-like state
corresponds to a highly excited states and have no analog in solid-state
systems. They can be realized by a sudden quench the effective 1D
interaction from the strongly repulsive to the strongly attractive
interaction regime by adjusting the magnetism field. The TG gas-like state can transfer into the STG gaslike
phase which are hard to realize in traditional condensed-matter physics \cite%
{Haller04092009, PhysRevA.81.031609, PhysRevA.83.053632}. The existence of
these stable gas-like states against cluster-like states due to the
existence of large Fermi-pressure-like kinetic energy inheriting from the
strongly repulsive interaction. Understanding the dynamic properties of the new synthetic bosonic
gas phase is an important subject.

As is well known, the mean-field theory typically does not work well for a
1D system, except in the very weakly interacting regime. The enhanced
quantum fluctuation is significant in 1D quantum systems \cite%
{PhysRev.130.1605, PhysRevLett.85.3745, PhysRevLett.86.5413,
PhysRevA.68.063605} which exhibit fascinating phenomena significantly
different from their three-dimensional counterparts. Thus when the
interaction is strong, non-perturbative methods such as the Bose-Fermi
mapping \cite{PhysRevLett.99.230402} or the Bethe ansatz \cite%
{PhysRevLett.20.98, 0295-5075-61-3-368} needs to be used to characterize the
features of the system properly. In the thermodynamic limit of $N,
L\rightarrow\infty$ and $N/L$ remains finite, the energy density and chemical potential of Bose gas
in TG and STG phase have been extracted from the Bethe-ansatz solution in
the absence of the external potential \cite{PhysRevA.80.043608,
PhysRevA.81.031609}. With the local-density approximation (LDA), a modified nonlinear Schr\"{o}dinger equation is
obtained. Using this kind of
nonlinear Schr\"{o}dinger equation, it is possible to investigate further the
dynamic properties of Bose gas in strong interactive regimes.

Associated with the periodicity and nonlinearity, there exist two important
waves in nonlinear periodical systems, namely Bloch waves and gap solitons
(GSs). Bloch waves, which exist in both linear and nonlinear periodic
systems, are extensive and spread over the whole space \cite{Aschcoft}. On
the contrary, GSs, which are spatially localized atomic wave packets, exist
only in a nonlinear periodic system \cite{book}. In particular, a class of
solitons called the fundamental gap solitons, have the major peak well
localized within a unit cell \cite{Louis}. The solitons with two peaks of
opposite signs within a unit cell are called the subfundamental gap solitons
\cite{PhysRevA.74.033616}. The existence and stability of GSs are important
issues. The relationship between the GSs and the nonlinear Bloch waves
(NLBWs) is also a topic of considerable interest.

We have studied the GSs and NLBWs of interacting bosons in
one-dimensional optical lattices, taking into account the repulsive
interaction from the weak to the strong limits \cite{PhysRevA.83.043610}.
The composition relation between the GSs and NLBWs was verified numerically to exist for
the whole span of the interaction strength. The stable GSs was found to form easily in a weakly
interacting system with energies near the bottom of the lower-level linear
Bloch band gaps. The present issues are whether the stable GSs can exist in
the stable STG phase, and whether the composition relation remains valid?
To what extent the GSs and NLBWs change when the interaction changes from
repulsive to attractive case?

In this paper, we attempt to investigate the GSs and NLBWs of the so-called STG
gas in a 1D optical lattices. We are interested particularly in its
existences and stability. It will be shown that the amplitude of periodical
potential and nonlinear interactions are two important factors for the stability
of GSs. The linear stability analysis indicates that stable gap soliton
waves is easy to form near the bottoms of the linear Bloch band gaps. The
composition relation remains valid in STG phase.

The paper is organized as follows. In Sec.~\ref{Model equation}, we
introduce the model equation for a 1D periodic Bose system in STG phase, and
then present the Gaussian-like Bose density profile $\rho(x)$ within a unit
cell and chemical potential for different interaction constant $|c|$ in
order to understand the special Bose system from an alternating perspective. In Sec.~\ref{Gap
solitons and their stabilities}, theoretical analysis and numerical
simulations are used to investigate the stabilities of different family GSs
upon the changes of the interaction and the strength of the periodic
potential. We show, in Sec.~\ref{Composition relationship}, that the
composition relation exists between the NLBWs and GSs in the present case. A
generalized composition relation between the high-order solitons and
multiple periodic waves is also shown in this section. Sec.~\ref{Summary} is
a brief summary.

\section{Model equation}

\label{Model equation}

We consider a 1D periodic Bose system described by the following modified
nonlinear Schr\"{o}dinger equation
\begin{equation}
\left[ -\frac{\hbar ^{2}}{2m}\frac{d^{2}}{dx^{2}}+V_{ext}(x)+\tilde{F}\left(
\rho \right) \right] \Phi (x)=\mu \Phi (x),  \label{modified NSE}
\end{equation}%
where $V_{ext}(x)=v\cos (\frac{2\pi }{\Lambda }x)$ is the periodic potential
with $\Lambda $ the lattice constant and $v$ the strength. $\tilde{F}\left(
\rho \right) $ is responsible for the interaction energy with $\rho =|\Phi
|^{2}$. The nonlinear Schr\"{o}dinger equation (\ref{modified NSE}) is
obtained by a minimization of the free-energy functional $\mathcal{F}=%
\mathcal{E}-\mu N$. The chemical potential $\mu $ is introduced as a
Lagrange multiplier and $N$ is particle number. The energy functional $%
\mathcal{E}$ can be represented as
\begin{equation*}
\mathcal{E}=\int dx\left[ \Phi ^{\ast }\left( -\frac{\hbar ^{2}}{2m}\frac{%
d^{2}}{dx^{2}}+V_{ext}\right) \Phi +\rho \epsilon \left( \rho \right) \right]
\end{equation*}%
in the LDA where the system is assumed in local equilibrium at each point $x$
in the external trap. The first gradient term represents additional "local"
kinetic energy. The second term is considered as external potential energy.
The ground-state energy density can be expressed as $\epsilon \left(\rho
\right) =\frac{\hbar ^{2}}{2m}\rho ^{2}e\left( \gamma \right) $. So $\tilde{F%
}\left( \rho \right) =\frac{\partial }{\partial \rho }\left[ \rho \epsilon
\left( \rho \right) \right] $ with the normalization condition $\int dx|\Phi
(x)|^{2}=N$. The quantum
\begin{equation}
e\left( \gamma \right) =\frac{4\pi ^{2}}{3}\frac{1+p_{1}|\gamma
|+p_{2}\gamma ^{2}+p_{3}|\gamma |^{3}/4}{1+q_{1}|\gamma |+q_{2}\gamma
^{2}+p|\gamma |^{3}}  \label{e_gamma}
\end{equation}%
which is obtained by the Bethe-ansatz technique in the attractive STG phase
\cite{PhysRevA.81.031609}. Here $\gamma \equiv {c}/{\rho }$ with $c\equiv
mg/\hslash ^{2}$ ($g$ is the scattering length) and $p_{1}=0.075$, $%
p_{2}=0.013$, $q_{1}=0.227$, $q_{2}=0.034$, and $p=0.004$ are the fitting
parameters. Contrary to the case in repulsive interaction where $c>0$, here $
c<0$ in the regime from the weak- to the strong attractive interaction of
STG phase.

For convenience, dimensionless scaling will be made for the length and the
energy. Position $x$ is to be scaled in the unit of $\Lambda /(2\pi )$.
Periodic potential $V_{ext}(x)$, interaction energy $\tilde{F}\left( \rho
\right) $, and the chemical potential $\mu $ are all scaled in the unit of $%
8E_{r}$ with $E_{r}=\hbar ^{2}\pi ^{2}/2m\Lambda ^{2}$ being the recoil
energy. We obtain the following dimensionless time-independent nonlinear
Schr\"{o}dinger equation
\begin{equation}
\left[ -\frac{1}{2}\frac{d^{2}}{dx^{2}}+v_{0}\cos (x)+F\left( \rho \right) %
\right] \Phi (x)=\mu \Phi (x).  \label{reduced_NSE}
\end{equation}%
Eq.~(\ref{reduced_NSE}) is the starting point of the calculations throughout
this paper.

\begin{figure}[t]
\includegraphics[width=8cm]{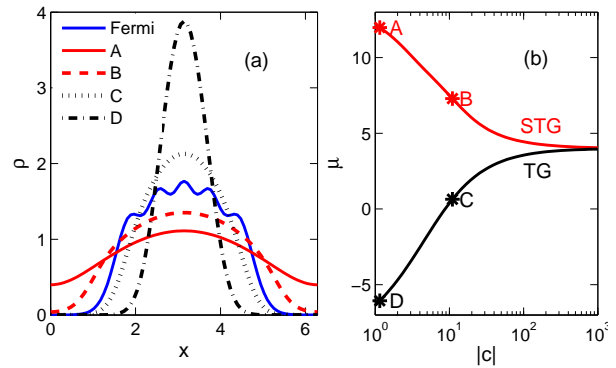}
\caption{(Color online) Panel (a): The Gaussian-like Bose density profile $%
\protect\rho(x)$ within a unit cell for different interaction constant $|c|$%
. For the repulsive case, $F(\protect\rho)$ is taken from the Ref.
\protect\cite{PhysRevA.80.043608}. The density profile labeled by Fermi is
obtained by the Bose-Fermi mapping method \protect\cite{PhysRevA.83.043610}.
The particle number is taken to be 50 and the periodic potential strength is
$v_0=10$. (b) The interaction constant $|c|$ dependence of chemical
potential. The marks A, B, C and D are used to show the parameters which have been used in panel (a).}
\label{fig1}
\end{figure}

In order to understand Bose gas in the STG phase intuitionally, we first
compare it to the repulsive interaction case. By solving Eq.~(\ref%
{reduced_NSE}) numerically to obtain the ground-state density profile $%
\rho(x)$ and the corresponding chemical potential $\mu$ for different
interaction constant $|c|$. It should note that the transition probability
from TG gas to STG phase is low for the weakly attractive regime \cite%
{Haller04092009, PhysRevA.81.031609}. In solving Eq.~(\ref{reduced_NSE}), we
first differentiate it using the finite-element method along with the
periodic boundary condition \cite{FEM}, and then evaluate several hundreds
of steps in imaginary time until the lowest chemical potential $\mu$ is
reached. The wave function is then obtained. In fact, the wave function obtained with this method is the NLBWs corresponding to the lowest chemical potential $\mu$. In the
calculation, the system is taken to be 10 lattice constant long ($x$ ranges
from $-10\pi$ to $10\pi$), the strength of periodic potential $v=10$, and
the particle number $N=50$ (average 5 particles per unit cell).

A Gaussian-like Bose density profile $\rho(x)$ within a unit cell is shown
in Fig.~\ref{fig1}(a). When $|c|$ is increased (see, for example, the case
of A and B points shown in Fig.~\ref{fig1}(b)), density profile increases at
the center while decreases at the two sides. It indicates that Bose gas in
STG phase has a larger effective (or equivalent) repulsive interaction in
weak attractive regime than that in strong attractive regime if we observe the properties of STG gas only from Eq. \ref{reduced_NSE} and Fig. \ref{fig1}. The existence of large effective (or equivalent)
repulsive interaction is, in fact, due to the existence of large kinetic
energy inheriting from the strongly repulsive interaction. The
repulsive interaction decreases with the increasing of $|c|$ which is
contrast to the case in TG phase (see, for example, the case of C and D). In
the extremely strong interaction regime ($|c|\rightarrow \infty$), the
chemical potential $\mu$ of the two phase will coincide in Fig.~\ref{fig1}%
(b) (See also the case B and C approaching the case Fermi in Fig.~\ref{fig1}%
(a)). In such limit, the system will behave similar to the noninteracting
fermions. We have used the Bose-Fermi mapping method to calculate the density profile.
The five-peak shell structure is shown in Fig.~\ref{fig1}(a) (blue solid line). When the large particle number is loaded in the system, a smooth density distribution will emerge.

\section{Gap solitons and their stabilities}

\label{Gap solitons and their stabilities}

\subsection{The general features of Gap solitons}

\begin{figure}[t]
\begin{center}
\includegraphics[width=8cm]{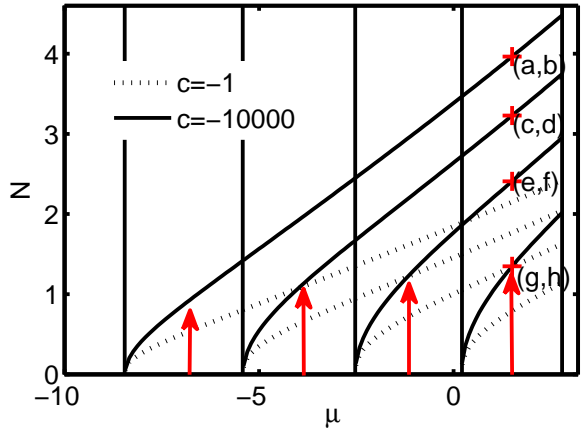}
\end{center}
\par
\vspace{-0.5cm}
\caption{(Color online) Particle number $N$ of FGSs as the function of
chemical potential $\protect\mu $ for different interaction constant $c=-1$ (dotted line),
$-10000$ (solid line) respectively. The points marked by red ``$+$" sign locate at the
center of the fourth band gap and will be studied in Fig.~\protect\ref{fig3}%
. The other parameters are same as that in Fig. \protect\ref{fig1}. The
arrows indicate the particle number contained in the GSs where the chemical potentials of the GSs are in the center of corresponding band gap.}
\label{fig2}
\end{figure}

We set $F(\rho )=0$ to solve the linear Schr\"{o}dinger equation (\ref%
{reduced_NSE}) exactly by the finite-element method. The periodic potential
strength $v_0$-dependent lowest four bands and band gaps are shown in Fig.~%
\ref{fig2}. A large $v_0$ is used in our calculations due to the large
nonlinear interaction of Bose gas in STG phase. So the Bloch bands reduce to
highly-degenerate thin levels. We then retain $F(\rho)$ to solve the
nonlinear Schr\"{o}dinger equation (\ref{reduced_NSE}) numerically. The
numerical solutions of GSs are obtained by differentiating Eq.~(\ref%
{reduced_NSE}) on a finite difference grid to obtain a coupled algebraic
equations and solve it with the Newton-relaxation method \cite%
{PhysRevA.74.033616}.

Figure \ref{fig2} shows particle number $N$ of GSs as the function of $\mu$
for two different interaction constant $c$'s. As shown, for example, when $%
c=-10000$ close to the strong limit (solid lines), the first nonlinear Bloch
band (NLBB) develops from the first linear Bloch band (LBB) and is lifted
with the increase of $N$. The high-order NLBB develops analogously from the
high-order LBB and is lifted with the increase of $N$. The so-called NLBB
lifting is simply due to the fact that the larger $N$ is, the larger the
nonlinearity and hence the corresponding $\mu$ are. For weaker interaction
case, $c=-1$ (dotted lines), the NLBB is seen to correspond to less $N$
for the same $\mu$, as compared to those of the $c=-10000$ case. This is
because a large nonlinear interactions correspond to the smaller interaction constant case, one does'nt needs a larger $N$ to
achieve the same nonlinear effect.

\begin{figure}[t]
\includegraphics[width=8.5cm]{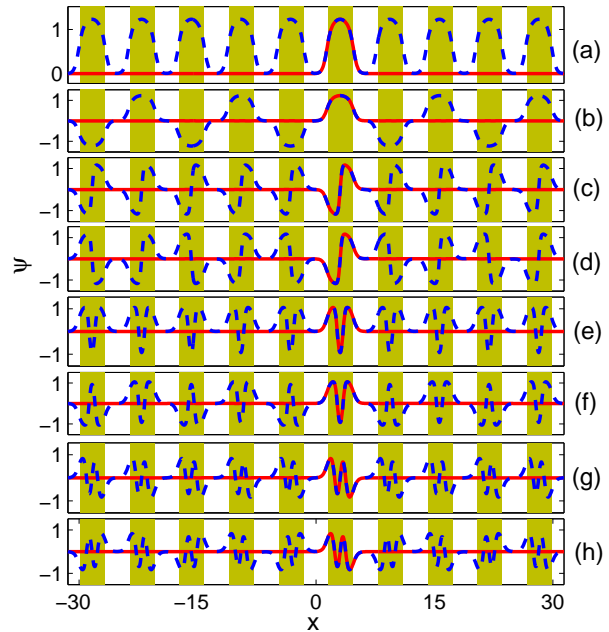}
\caption{(Color online) The different family GSs (red solid line) and NLBW
(blue dashed line) in the forth LBB gap marked in Fig. \protect\ref{fig2}.}
\label{fig3}
\end{figure}

As shown previously, NLBB can be viewed as the lifted LBB by increasing the
nonlinear interaction. While LBB can be viewed as the evolution from the
discrete energy levels of an individual well \cite{Zhang1, PhysRevA.83.043610}. Therefore it is not difficult
to think that NLBW belonging to the $n$th NLBB should have {$n$-$1$} nodes
(in the sense of the $n$th bound state) in an individual well of the
periodic potential. Therefore, GSs should behave like the bound states in an individual
well of the periodic potential in a sense. The different family GSs are presented in
Fig. \ref{fig3}. The chemical potentials are all taken in the center of the
fourth band gap to show the GSs completely. The GS waves in Fig. \ref{fig3}
(a) and (b) belong to the first family and hence no node. The difference of
NLBWs between Fig. \ref{fig3} (a) and (b) is that the later obtain a $\pi$%
-phase difference in adjacent well. The GS waves in Fig. \ref{fig3} (c) and
(d), however, belong to the second family and hence one node. Therefore, the
above conclusion remains correct when the nonlinear term is replaced by the $%
F(\rho)$ of Bose gap in STG phase.

\subsection{The stabilities of different family GSs}

We shall study the linear stability of various GS solutions in this subsection. As we know, the
bright solitons form for atomic matter waves when the linear spreading due
to kinetic energy is compensated by the attractive interaction between
atoms. Similarly, the existence of GSs is due to that the linear spreading
and repulsive atom-atom interaction are compensated by the confinement of
periodical potential. It is therefore interesting to study the effect of the
nonlinear and periodical potential on the GS stability. Following the
standard procedure in Ref. \cite{PhysRevA.83.043610}, we first add a small
perturbation $\Delta\Phi(x,t)$ to a known solution $\Phi(x)$, and then
insert the perturbation into Eq.~(\ref{reduced_NSE}). One then obtains the
linear eigen equations by dropping the higher-order terms. Among all
eigenvalues, if there exists a finite imaginary part, the solution of
$\Phi (x)$ would be unstable. Otherwise, the solution of $\Phi(x)$ is stable.

\begin{figure}[t]
\vspace{0.2cm} \includegraphics[width=8cm]{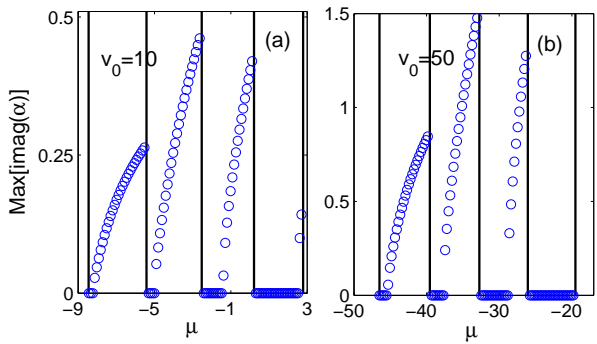}
\caption{(Color online) Studies of the stability of the FGSs for different
families. The interaction
constant $c=-10000$. The periodical potentials are taken (a) $v=10$ and (b) $v=50$ respectively. The other parameters are the same as that used in Fig.
\protect\ref{fig1}.}
\label{fig4}
\end{figure}

The stability of GSs is investigated in Fig.~\ref{fig4} for different
amplitude of periodical potential $v_0=10$ and $v_0=50$. It indicates that
the first family GSs which develop from the first LBB are stable when their
chemical potentials $\mu$ are near the bottom of the first LBB gap. They
will become unstable when $\mu$ becomes higher within the first band gap
(with the increase of $N$), and enters into the second and third band gaps
(not shown in Fig.~\ref{fig4}). Similarly, the second and high family
GSs are also stable when the chemical potentials are near the bottom of the
corresponding band gap.

The above behaviors are very similar to the repulsive case. This is due to
the little particle number contained in GSs (See Fig. \ref{fig2}) while the
GSs develop from the bottoms of band gaps. In such case, the particles
behavior is similar to the non-interaction particles in a single potential
well. In such case, the GSs are, in fact, the stationary states in discrete energy
levels and therefore stable. When increasing the particle number, the
interaction between particle will increase. If the interaction energy cannot
be compensated by the confinement potential energy when the confinement
potential remains unchanged, the GSs will become unstable. Seen also from
Fig. \ref{fig3}, the wave functions of high-order family GS have more nodes. It is therefore that the kinetic energy of GS is larger in high-order family than that of in low-order family. From the above
analysis, it also indicates the interaction is the main factors to govern the GS satiability.

\begin{figure}[t]
\includegraphics[width=7cm]{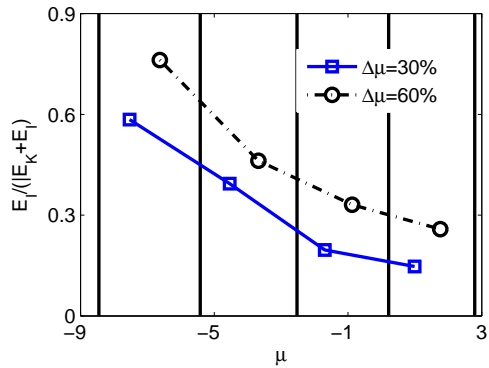}
\caption{(Color online) Shown $\protect\eta(\protect\mu)=E_I/(E_K+E_I)$ for
different family GSs. The chemical potentials $\protect\mu$ are taken $0.3
$ (blue solid line marked with square) and $0.6$ (black dashdot line marked
with circle) times band gap width away from the bottom bands respectively. The
other parameters are the same as that used in Fig. \protect\ref{fig4} (a).}
\label{fig5}
\end{figure}

It is interesting to see that the first family GSs are stabile only
in a narrow regimes near the bottom of the first bang gap. This regime
becomes wide for high family GSs. See the length of red arrows in Fig. \ref{fig2},
the particle number contained in GSs are equal approximately in the center of band gaps. If judging the GS stability according to the general intuition, the high family GSs should be unstable due to large kinetic energy.
In order to understand the GS stability
in different family quantitatively, we multiply $\Phi(x)$ in the left side
of Eq. \ref{reduced_NSE} and integrate in whole space to obtain the kinetic
energy $E_K=\int dx |\frac{\partial\Phi(x)}{\partial x}|^2$ and interaction
energy between atom $E_I=\int dx F(\rho)\rho$. The quantum $%
\eta(\mu)=E_I/(E_K+E_I)$ is defined to indicate the effect of repulsive
interaction between atoms. The number results of $\eta(\mu)$ are presented
in Fig. \ref{fig5}. The chemical potentials $\mu$ are taken $0.3$ and $0.6$
times width of the band gap away from the bottom bands respectively. It is clearly that
the repulsive interaction of high family GSs becomes weak relative to the the kinetic energy. Therefore, even though with the same particle
number, the low-order family GSs are tend to unstable and the high-order family GSs
is expected to be stale however.

\subsection{The effects of periodical potential}

Comparing the linear stability analysis presented in Fig. \ref{fig4} (a) and
(b), the fourth family GSs are only unstable in a narrow regime near
the top of band gap when $v_0=10$ (Fig. \ref{fig4} (a)). However, the GSs
are all stable in the entire regime of the fourth band gap when $v_0=50$
(Fig. \ref{fig4} (b)). The increasing of stable regime seems not to be
obvious for low-order family GSs when increasing the amplitude of periodical
potential.

When increasing periodical potential, the unstable GSs will become stable
GSs since the confinement of periodical potential is enough to compensate
the linear spreading and repulsive atom-atom interaction. So the stable
regime near the bottom of band gap will become wide in such as case. As the
above discussions, the effect of nonlinear interaction for high-order family
GSs are weaker than that of low-order family GSs. It results the variations of the
stable regime is obvious for high-order family GSs when increasing the
periodical potential. We have also studied the stability of Bose near TG
phase numerically. The above behaviors are still right (not present here).

\section{Composition relationship}

\label{Composition relationship}

As shown previously, the different family GSs originate from the stable
bound state of single periodical well and develop in the band gaps. On the other hand, NLBB
can be viewed as the lifted LBB by increasing the nonlinear interaction.
However, LBB can be viewed as the evolution from the discrete energy levels of
an individual well. Therefore GSs and NLBWs should have some similarity. It has shown that GSs and NLBWs belonging to the $n$th NLBB have {%
$n$-$1$} nodes in an individual well of the periodic potential. In
particular, it has also pointed out and proven numerically in the whole
interaction regime that GSs are the building blocks of the NLBWs \cite%
{Zhang1, PhysRevA.83.043610}. The current issue is whether the above
statements remain correct when the nonlinear term is replaced by the case
of Bose atoms in the STG phase.

In Fig.~\ref{fig3}, we have plotted both NLBW (blue dashed line) and GSs
(red solid line) for Bose in STG phase ($c=-10000$). An almost perfect
(unnoticeable) match is found between the NLBW and the GS within one unit
cell. The good match occurs regardless of the first family or the high
family GSs. It thus gives a strong evidence that GSs can be considered as
the building blocks of the NLBWs. It is important to note that the waves
shown in Fig.~\ref{fig3}(a)\&(b) are belonging to the first-family GS and
hence of no node. In Fig.~\ref{fig3}(c)\&(d) however, GS belong to the
second-family and hence of one node. The good matches are also found between
the NLBW and the GS of high-order family in Fig.~\ref{fig3}(e)\&(f) and (g)\&(h) where multiply notes exist within
one unit cell.

\begin{figure}[t]
\begin{center}
\includegraphics[width=8.5cm]{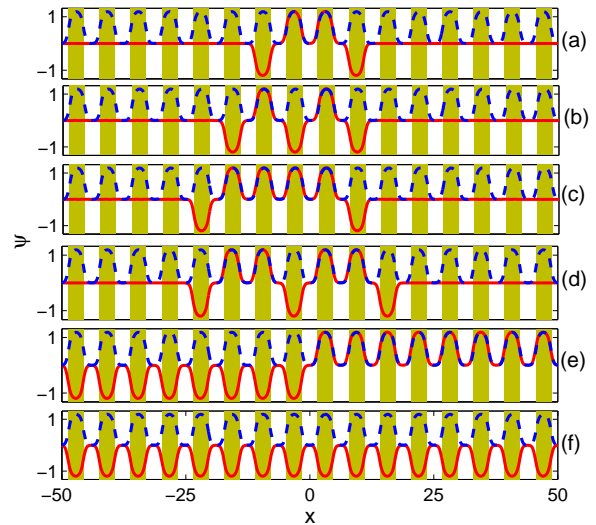}
\end{center}
\caption{(Color online) Illustration of the high-order gap solitons (red
solid line) and NLBWs (blue dashed line) for $c=-10000$ and $\protect\mu =0.20
$. The periodic potential strength is $v=10$ for all panels.}
\label{fig6}
\end{figure}

In addition to the GSs and NLBWs studied previously, there are two other
types of waves which are also common in a nonlinear system. One is called
the high-order GSs which are GS waves of multiple peaks over multiple unit
cells \cite{Wang}. Another is called the multiple periodic waves which are
defined as $\Phi(x)=\exp(ikx)\psi_{k}(x)$ with $\psi_{k}(x)=\psi_{k}(x+2n\pi)
$ and $n$ being a positive integer \cite{PhysRevA.69.043604}. In the
repulsive atom interaction, it has been shown that composition relation
between the GSs and the NLBWs can be generalized to construct multiple
periodic waves from the high-order GSs \cite{Zhang2, PhysRevA.83.043610}.
The present issue is again to see whether the generalized composition
relation remains valid in the present case of Bose gas in STG phase.

We have first solved the nonlinear Schr\"{o}dinger equation (\ref%
{reduced_NSE}) numerically using the imaginary-time method with the periodic
boundary conditions. The NLBW function $\Phi(x)$ and chemical potential $%
\mu=0.20$ are obtained accurately with a given interaction constant $c=-10000$%
. The Newton-relaxation method is then used to solve the nonlinear Schr\"{o}%
dinger equation (\ref{reduced_NSE}) to obtain high-order GSs with the
pre-obtained $\mu=0.20$ as done in the case of Fig. \ref{fig3} with the
proper guess wave function.

Fig.~\ref{fig6} (a) show the four-peak high-order GSs (two center peaks up,
two adjacent peaks down). An almost perfect match is found between the
high-order GSs and the NLBWs except the phase of wave function in some
single well. Thus high-order GSs can also be viewed as the truncated NLBWs
in the whole interaction regimes. The Newton-relaxation method is further
used to solve the nonlinear Schr\"{o}dinger equation (\ref{reduced_NSE}) to
obtain the multiple periodic waves in Fig.~\ref{fig6} (b) with the
pre-obtained $\mu=0.20$. Five periodic high-order GSs in Fig.~\ref{fig6} (b)
has been used as the initial guess wave function. An almost perfect match is
found between the multiple periodic waves and the NLBWs except the phase of
wave function in some single well. We also present the multiple periodic
waves in Fig.~\ref{fig6} (c) (five-peak periodic waves) and Fig.~\ref{fig6}
(d) (ten-peak periodic waves). In particular, we presents twenty-peak
periodic waves in Fig.~\ref{fig6} (e). The best matches between GSs and
high-order GSs, as well as multiple periodic waves and NLBWs, strong support
that the GSs are the basic entity and the other waves can built by them.

The above conclusions are useful to obtain NLBWs and multiple periodic waves
of nonlinear Schr\"{o}dinger equation. In actual numerical calculation, we
can first solve the nonlinear Schr\"{o}dinger equation to obtain GSs of
different family, and then build an initial wave function with the GS. If we
want to obtain $n$-periodic waves in $m$-periodic potential well, the
initial guess wave function can be built by $n$-GSs as a block ($m/n$ is
assumed to be a integer). If we want to obtain NLBWs in $m$-periodic
potential well, the initial guess wave function can be built by $m$-GSs
arrangement one by one. With the initial wave function, the
Newton-relaxation method is again used to solve the nonlinear Schr\"{o}%
dinger equation (\ref{reduced_NSE}) to obtain GSs and multiple periodic
waves accurately. The advantage of this method is that the initial wave function is very close to the finial wave function. The iteration times decrease obviously.

\section{Summary}

\label{Summary}

In summary, we have investigated the GSs of 1D periodic bosonic gas in STG
phase. The main focus is to consider its stabilities. By the linear
stability analysis, it is found that the periodic potential and the
nonlinear interactions are important to the stabilities of GSs. Increasing
the amplitude of periodic potential or decreasing the nonlinear
interactions, the unstable GSs can turn into stable. It is particular that
the high family of GS is easy to form near the bottoms of the LBB
gaps. Our numerical results further verify that the composition relation between various GSs and NLBWs does exist generally. It gives an
alternative way to obtain the NLBWs. It is worth emphasizing that the above
conclusions are also valid to Bose gas in the whole repulsive
interaction regime.

\begin{acknowledgments}
This work is supported by Hebei Provincial Natural Science Foundation of China (Grant
No. A2010001116), and the National Natural Science
Foundation of China (Grant No.s 10974169, 11174115, 10934008, 41174116).
\end{acknowledgments}


%

\end{document}